\documentclass[conference]{IEEEtran}
\IEEEoverridecommandlockouts
\usepackage{cite}
\usepackage{amsmath,amssymb,amsfonts}
\usepackage{algorithmic}
\usepackage{algorithm}
\usepackage{graphicx}
\usepackage{verbatim}
\usepackage{textcomp}
\usepackage{xcolor}
\usepackage{url}
\usepackage{bm}
\usepackage{subfigure}
\usepackage{flushend}
\usepackage{tabularx}
\usepackage[
    top=0.73in,
    bottom=0.99in,
    left=0.7in,
    right=0.7in,
]{geometry}

\def\BibTeX{{\rm B\kern-.05em{\sc i\kern-.025em b}\kern-.08em
    T\kern-.1667em\lower.7ex\hbox{E}\kern-.125emX}}
\begin{document}

\title {
A Privacy-Preserving Federated Learning Method\\ 
for Heterogeneous Clients
}

\author{
\IEEEauthorblockN{Yusaku Negoya\IEEEauthorrefmark{1}, Feifei Cui\IEEEauthorrefmark{2}, Zilong Zhang\IEEEauthorrefmark{2}, Miao Pan\IEEEauthorrefmark{3}, Tomoaki Ohtsuki\IEEEauthorrefmark{4}, and Aohan  Li\IEEEauthorrefmark{1}
} 
\IEEEauthorrefmark{1}Department of Computer and Network Engineering, The University of Electro-Communications, Tokyo, Japan\\
\IEEEauthorrefmark{2}School of Computer Science and Technology, Hainan University, Haikou 570228, China.\\
\IEEEauthorrefmark{3}Department of Electrical and Computer Engineering, University of Houston, Houston, TX, 77204, USA\\
\IEEEauthorrefmark{4}Department of Information and Computer Science, Keio University, Yokohama, Japan\\
}
\maketitle

\begin{abstract}
Federated Learning (FL) with Differential
Privacy (DP) can ensure the protection of healthcare data privacy
against malicious user attacks. However, FL with the DP method
faces an inherent trade-off: stronger privacy protection degrades
predictive accuracy due to injected noise. On the other hand, 
Homomorphic Encryption (HE) enables computations on encrypted
data and the aggregation of encrypted gradients without
introducing DP-induced noise, thereby potentially increasing the predictive accuracy.
However, it may increase the computation cost. To improve
the predictive accuracy while considering the computational
ability of heterogeneous clients, we propose a Privacy-Preserving Machine
Learning (PPML)-Hybrid method. 
In the proposed PPML-Hybrid method, clients
distributedly select either HE or DP based on their computational
resources, so that HE clients contribute noise-free updates while
DP clients reduce computational overhead. Meanwhile, clients with high computational resources can flexibly adopt HE or DP according to their privacy needs. Performance evaluation on omics datasets shows that our proposed method achieves
comparable predictive accuracy while significantly reducing computation time relative to HE-only. Additionally, it outperforms
DP-only methods under equivalent or stricter privacy budgets.

\end{abstract}


\section{Introduction}
\label{sect:introduction}

Artificial Intelligence (AI), particularly deep learning, enables integrative analysis of multi-omics data (genomics, transcriptomics, proteomics) to extract complex biological patterns \cite{Zhou2024_AutoBA}. 
However, omics data contain highly sensitive personal information, creating significant risks of privacy leakage from learned models \cite{Kaissis2020}. 
Thus, especially in multi-institutional collaborations, strict confidentiality is ethically and technically essential to ensure trustworthy deployment. 
Federated Learning (FL), a framework that enables collaborative training of machine learning models locally without sharing raw data, has emerged as a promising solution to this privacy concern \cite{mcmahan2017communication}. 
Nevertheless, FL alone does not provide complete privacy guarantees against sophisticated attacks, such as model inversion attacks and membership inference attacks \cite{bertran2023scalable}.


To address this vulnerability, applying Differential Privacy (DP) to FL has been extensively studied \cite{pihur2022ldp}. DP provides a strong, mathematically rigorous framework for privacy by adding carefully calibrated noise to the computation, thereby stochastically obscuring the contribution of any single individual. 
The prior work, Privacy Preserving Machine Learning in Omics data (PPML-Omics), extended this approach by combining DP with Decentralized Randomization (DR) \cite{zhou2024ppml}. This method anonymizes gradient information by shuffling it among clients, which mitigates accuracy degradation with an increasing number of clients and protects privacy without relying on a trusted third party. 
However, as its privacy foundation rests on DP, PPML-Omics cannot fully escape the fundamental trade-off between privacy and utility, where a stronger privacy guarantee (i.e., more noise) inevitably leads to a decline in model accuracy. 

As an alternative that preserves model accuracy, Homomorphic Encryption (HE) allows computations on encrypted data and enables aggregation of encrypted gradients without DP-induced noise \cite{he2025}. In practice, however, HE introduces substantial computational overhead due to expensive encryption/decryption and ciphertext-domain operations\cite{zhang2024heprune}.
These factors slow down iterative training and limit the applicability of FL to large-scale settings involving complex deep models and numerous clients. 
Moreover, recent studies have investigated hybrid privacy-preserving approaches that combine HE and DP, in which each client employs both techniques to protect privacy \cite{grivet2023combining}. 
However, the heterogeneity of computational resources among clients is not considered.

To address the respective challenges of DP (lightweight but potentially lossy) and HE (high-fidelity but costly), we propose a hybrid approach termed as PPML-Hybrid. 
Rather than fixing roles centrally, clients autonomously select either DP or HE based on their computational resources. In effect, some clients contribute noise-free HE updates while others adopt DP to reduce overhead, yielding an asymmetric mix that balances privacy, utility, and training time. The contributions of this paper are summarized as follows.

\begin{itemize}
    \item To improve model accuracy while considering both computational ability and potential security requirements of the clients, we propose a PPML-Hybrid method, in which clients with high computational resources can flexibly select either DP method or HE method, whereas resource-limited clients adopt DP method.

    \item We evaluate our proposed method with omics data and observe its superiority in the privacy–utility–efficiency trade-off compared to DP-only and HE-only baselines. Moreover, we evaluate the ratio of DP-based to HE-based clients, showing that increasing the proportion of HE-based clients improves utility but increases computational time, thereby highlighting the trade-off among privacy, utility, and efficiency.

    
\end{itemize}


This paper is organized as follows. Section II reviews preliminary knowledge. Section III describes the proposed PPML-Hybrid method. Section IV presents the performance evaluation. Section V concludes the paper.

\section{Preliminary}
\label{sect:preliminary}
In this section, we provide an overview of the foundational privacy-preserving technologies and the specific prior work that our proposed method builds upon, incorporating their privacy requirements.

\subsection{Differential Privacy}
DP offers a mathematically rigorous framework for privacy protection by ensuring that the output of a data analysis process is not significantly influenced by the presence or absence of any single individual's data in the dataset.
Specifically, a randomized algorithm $\mathcal{M}$ satisfies $(\epsilon, \delta)$-DP if, for any adjacent datasets $D$ and $D'$ (differing by at most one record) and for any set of possible outputs $S$, the following inequality holds:
\begin{equation}
    \mathbb{P}(\mathcal{M}(D) \in S) \le e^{\epsilon} \mathbb{P}(\mathcal{M}(D') \in S) + \delta. 
\end{equation}
Here, $\epsilon$ is the privacy budget, where a smaller value implies stronger privacy, and $\delta$ is a probabilistic relaxation term. In FL, DP is typically achieved by adding carefully calibrated noise to the gradients. However, this introduces a fundamental trade-off: stronger privacy guarantees (a smaller $\epsilon$) require more noise, which can degrade the model's utility and accuracy. 

\subsection{Homomorphic Encryption}
HE is a cryptographic technique that allows computations to be performed directly on encrypted data. Within this field, the Cheon-Kim-Kim-Song (CKKS) scheme is particularly relevant for machine learning applications, as it is designed to efficiently handle computations on approximate real numbers (i.e., floating-point numbers) \cite{yao2016ckks}.
Denoting the encryption function as $\text{Enc}(\cdot)$, the decryption function as $\text{Dec}(\cdot)$, and plaintexts as $m_1, m_2$, the additive homomorphic property of CKKS can be expressed as:
\begin{equation}
    \text{Dec}(\text{Enc}(m_1) \oplus \text{Enc}(m_2)) \approx m_1 + m_2,
\end{equation}
where $\oplus$ represents the addition operation in the ciphertext space. This enables a server to aggregate encrypted gradients from multiple clients without decrypting them, thus preserving privacy without compromising model accuracy. The primary drawback of HE, including CKKS, remains its substantial computational overhead for encryption, decryption, and ciphertext operations.

\subsection{PPML-Omics}

PPML-Omics is a prior work on a privacy-preserving FL method tailored for omics data. This method combines DP with DR, i.e., shuffling updated models among clients. Specifically, after each client adds DP-compliant noise to their gradient, the gradients are shuffled within the client population, thus anonymizing the origin of each gradient.
This shuffling process provides an effect of amplification of the privacy budget. As a result, the overall privacy guarantee ($\epsilon$-value) achieved by the system becomes stronger than that of standard DP-FL, even when the level of noise added by each client is the same. In other words, PPML-Omics can achieve an equivalent level of privacy with less noise, thereby mitigating model accuracy degradation while protecting privacy. However, since its privacy foundation still relies on DP, the inherent trade-off between privacy and utility has not been completely resolved. 

\section{Proposed Method}

\begin{figure}
    \centering
    \includegraphics[width=1\linewidth]{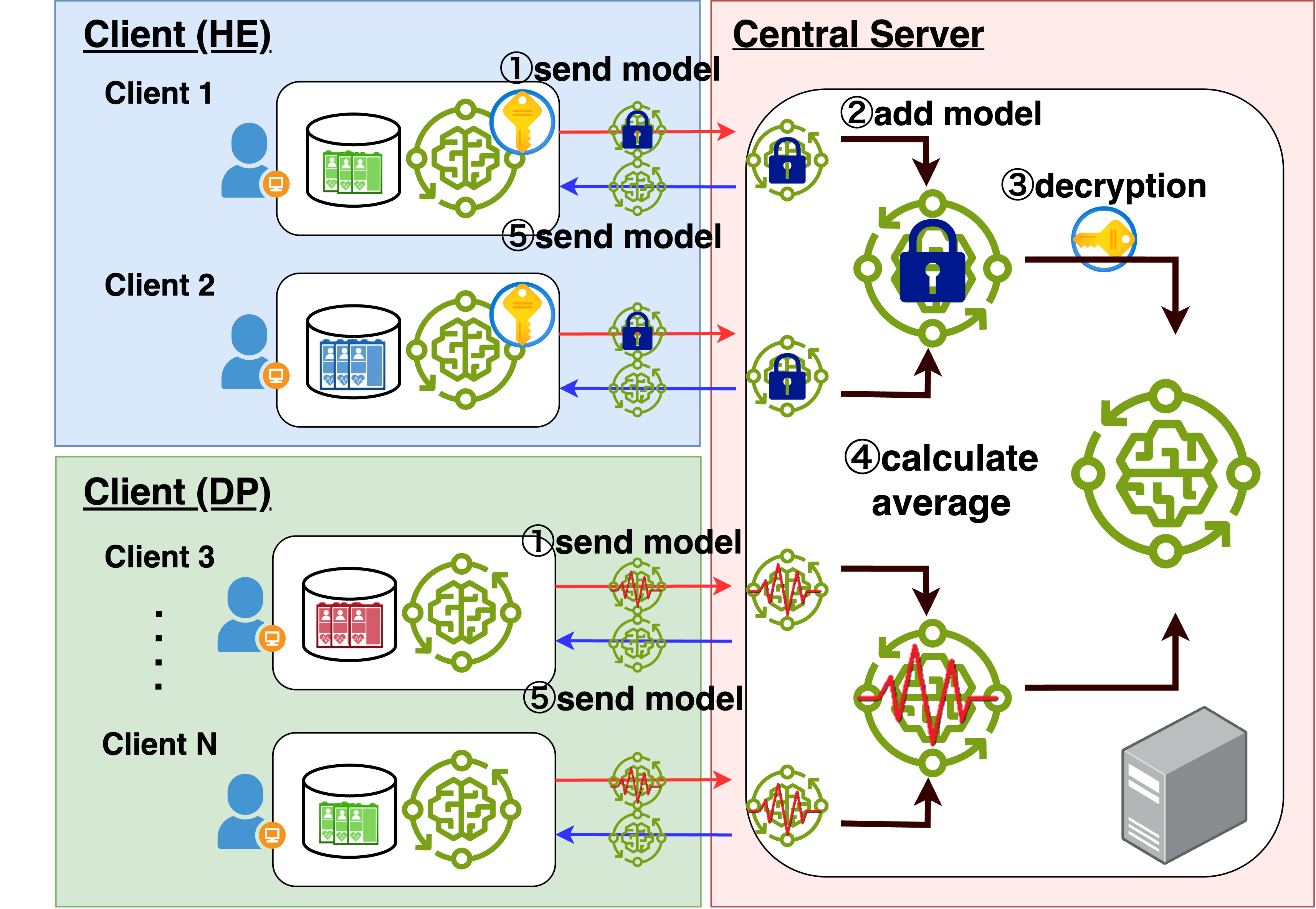}
    \caption{The proposed hybrid architecture with HE and DP clients.}
    \label{fig:hybrid_architecture}
\end{figure}

In this section, we present the architecture of our proposed hybrid privacy-preserving federated learning system and the detailed learning process.

\subsection{System Architecture}

The proposed system consists of a single central server and $N$ participating clients, as illustrated in Fig~\ref{fig:hybrid_architecture}. 
Each client autonomously selects its privacy mode (HE or DP) based on its privacy budget, privacy requirement, and available computational resources. It is important to note that this selection mechanism is currently proposed as a flexible framework. Since the specific criteria for HE/DP selection are beyond the scope of this paper, users are expected to adjust the parameters to suit their deployment environment manually. In practical scenarios, organizations with higher computational and security requirements may prefer HE to ensure the cryptographic confidentiality of updates. In contrast, clients with lower computational capacity or less stringent privacy requirements may adopt DP for lightweight local protection. This hybrid configuration enables dynamic optimization of privacy–utility–efficiency trade-offs across heterogeneous participants.
Let $\alpha$ denote the ratio of HE clients to the total number of participating clients.
\begin{itemize}
    \item \textbf{The HE Group ($N_{HE} \approx \alpha N$ clients):} Clients with sufficient computational resources leverage the CKKS homomorphic encryption scheme to protect their model updates.
    \item \textbf{The DP Group ($N_{DP} \approx (1-\alpha)N$ clients):} Clients with limited computational resources utilize a computationally lightweight DP protocol for protection. 
\end{itemize}

As shown in Fig~\ref{fig:hybrid_architecture}, a single training round can be briefly summarized as follows. After local training, each client sends its updated model to the server (\textcircled{\raisebox{-0.9pt}{\scalebox{0.8}{1}}}). The server then aggregates the updates from each group separately (\textcircled{\raisebox{-0.9pt}{\scalebox{0.8}{2}}}). For the HE group's aggregated update, the server performs decryption (\textcircled{\raisebox{-0.9pt}{\scalebox{0.8}{3}}}). Next, the server calculates the average of the aggregated updates from both groups to create the new global model (\textcircled{\raisebox{-0.9pt}{\scalebox{0.8}{4}}}). Finally, this updated global model is sent back to all participating clients for the next round of training (\textcircled{\raisebox{-0.9pt}{\scalebox{0.8}{5}}}).



\subsection{Algorithm and Learning Process}
\par
\vspace*{6pt}
\begin{algorithm}[!t]
\vspace{6pt}
\caption{PPML-Hybrid}
\label{alg:hybrid_fl}
\begin{algorithmic}[1]
\STATE Initialize global model weights $W_0$, set $t=0$, clipping norm $C$, learning rate $\eta$
\FOR{$t < T-1$}
    \STATE \# Clients self-select mode and join $C_{HE}$ or $C_{DP}$ based on local computational resources and privacy requirements.
    \FOR{each client $k \in \{C_{HE} \cup C_{DP}\}$ in parallel}
        \STATE $u_k \leftarrow \text{ClientUpdate}(W_t, D_k)$
    \ENDFOR
    \STATE \# Server aggregates updates from each group
    \STATE \# Aggregate from HE Group
    \STATE $\text{Enc}(\mathbf{G}_{HE}) \leftarrow \bigoplus_{k \in C_{HE}} u_k$
    \STATE $\mathbf{G}_{HE} \leftarrow \text{Decrypt}(\text{Enc}(\mathbf{G}_{HE}))$
    \STATE \# Aggregate from DP Group
    \STATE $\mathbf{G}_{DP} \leftarrow \sum_{k \in C_{DP}} u_k$
    \STATE \# Update the global model
    \STATE $W_{t+1}^{'} \leftarrow W_t^{'} - \eta \frac{\mathbf{G}_{HE} + \mathbf{G}_{DP}}{N}$
\ENDFOR
\STATE
\STATE \textbf{function} ClientUpdate($W_t, D_k$):
\STATE \# Compute gradient on local data using SGD
\STATE $\mathbf{g}_k \leftarrow \nabla L(W_t; D_k)$
\STATE \# Client-side autonomous mode selection (conceptual; no centralized rule is enforced)
\STATE $\text{mode}_k \leftarrow \text{SelectModeByResources}() \in \{\text{HE}, \text{DP}\}$
\IF{$\text{mode}_k == \text{HE}$}
    \STATE \textbf{return} $\text{Enc}(\mathbf{g}_k)$
\ELSE 
    \STATE $\tilde{\mathbf{g}}_k \leftarrow 
    \mathbf{g}_k \big/ \max\!\left(1, \frac{\|\mathbf{g}_k\|_2}{C}\right)
    + \mathcal{N}(0, \sigma^2 I)$
    \STATE \textbf{return} $\tilde{\mathbf{g}}_k$
\ENDIF
\end{algorithmic}
\end{algorithm}

A single round of our hybrid learning process, from distribution to aggregation, proceeds as follows:

\begin{enumerate}
    \item \textbf{Initialization:} The central server initializes the global model $W_0$ with a DenseNet-121 backbone pretrained on ImageNet, replaces the final layer with zero-initialized weights and a bias set to the training-data mean gene expression, and estimates dataset mean/std from the training set to apply Normalize(mean, std) in preprocessing.
    
    \item \textbf{Distribution:} In FL round $t$, the central server broadcasts the current global model weights $W_t$ to all clients.

    \item \textbf{Local Update:} Each client $k$ computes the gradient $\mathbf{g}_k = \nabla\mathcal{L}(W_t; D_k)$ on its local data. Using this gradient, the client then updates its local model by performing one step of the Stochastic Gradient Descent (SGD) algorithm.

    \begin{equation}
        W_t^k = W_t - \eta \mathbf{g}_k ,
    \end{equation}
    where $W_t^k$ denotes the updated local model weights for client $k$ after training on its local data $D_k$, and $\eta$ is the learning rate.

    The subsequent step depends on the client's self-selected mode:
    \begin{itemize}
        \item \textbf{HE Group:} The client encrypts its gradient $\mathbf{g}_k$ using an HE scheme (i.e., CKKS) and sends the ciphertext $\text{Enc}(\mathbf{g}_k)$ to the server.
        \item \textbf{DP Group:} The client applies $\ell_2$ clipping $\bar{\mathbf{g}}_k=\mathbf{g}_k/\max\!\bigl(1,\lVert\mathbf{g}_k\rVert_2/C\bigr)$, selects $\sigma$ via the analytic Gaussian mechanism \cite{Balle2018ImprovingTG}, adds noise to obtain $\tilde{\mathbf{g}}_k=\bar{\mathbf{g}}_k+\mathcal{N}(\mathbf{0},\sigma^2\mathbf{I})$ 
        for each client, and sends it to the server. Here, $C$ is the clipping norm, 
    \end{itemize}

    \item \textbf{Aggregation:} The server aggregates the received updates to compute the new global model $W_{t+1}$.
    \begin{itemize}
        \item It homomorphically adds all encrypted gradients from the HE group to obtain a single ciphertext representing their sum: $\text{Enc}(\mathbf{G}_{HE}) = \bigoplus_{k \in C_{HE}} \text{Enc}(\mathbf{g}_k)$. Here, $C_{HE}$ is the client set of the HE group.
        \item It decrypts this result to obtain the aggregate noise-free gradient for the HE group: $\mathbf{G}_{HE} = \text{Dec}(\text{Enc}(\mathbf{G}_{HE}))$.
        \item Concurrently, it sums all noisy gradients from the DP group: $\mathbf{G}_{DP} = \sum_{k \in C_{DP}} \tilde{\mathbf{g}}_k$. Here, $C_{DP}$ is the client set of the DP group.
        \item Finally, it combines the aggregates to update the global model:
        \begin{equation}
            W_{t+1}^{'} \leftarrow W_t^{'}- \eta \frac{\mathbf{G}_{HE} + \mathbf{G}_{DP}}{N},
        \end{equation}
        where $N$ is the number of participating clients in the round and $\eta$ is the learning rate.
    \end{itemize}
\end{enumerate}

This process is repeated until the model converges. The detailed procedure of our proposed algorithm, termed PPML-Hybrid, is summarized in Algorithm~\ref{alg:hybrid_fl}.

In line 1, the server initializes the global model weights $W_0$.
In line 3, clients autonomously decide their mode (HE or DP) based on local computational resources and thereby form $C_{HE}$ and $C_{DP}$ (no central enforcement of a fixed ratio).
In lines 4–6, each client calculates local updates $u_k$ in parallel using local data $D_k$.
In lines 8–10, the server decrypts the encrypted updates received from the HE group using homomorphic addition and decryption, and obtains the aggregate result $\mathbf{G}_{HE}$ of the HE group.
In line 12, the server sums the updates received from the DP group and obtains the aggregate result $\mathbf{G}_{DP}$ of the DP group.
In line 14, the server integrates the aggregated results of the HE group and the DP group to update the global model $W_{t+1}$.
In lines 17–27, the client-side local update procedure is defined. Each client first computes the gradient of the loss function on its local data using SGD. The client then self-selects HE or DP according to its available computational resources (conceptual policy), encrypting the gradient in the former case or adding calibrated noise in the latter.

\section{Performance Evaluation}
\label{sec:evaluation}

In this section, we evaluate the performance of the proposed PPML-Hybrid algorithm in terms of model accuracy, measured by the Mean Squared Error (MSE), and computational efficiency under varying numbers of clients. We first describe the experimental setup and parameters in detail. We then introduce the two baseline methods for comparison: a DP-only approach (PPML-Omics) and an HE-only approach. Finally, we present and discuss the experimental results in terms of model accuracy and computational time compared to the baselines.
\subsection{Experimental Setup}

\subsubsection{Dataset and Model}
We use a public spatial transcriptomics (ST) dataset of human breast cancer~\cite{stenbeck2021breastcancer} from 10x Genomics Visium, which was also utilized in PPML-Omics. The cohort comprises 23 patients and 68 H\&E-stained tissue sections, yielding 30{,}612 spatial spots in total. Each section is profiled on a regular Visium grid, where a spot has a diameter of approximately 100\,$\mu$m and a center-to-center spacing of approximately 200\,$\mu$m. For every spot, we extract a 224$\times$224\, px histology patch centered at the spot and pair it with the corresponding spot-level mRNA abundance vector. Across the dataset, 26{,}949 genes are detected. Following PPML-Omics, we perform a one-time gene selection at the beginning of the study (after library-size normalization and log transformation) and fix the evaluation to the 250 most highly expressed genes for all experiments; this keeps the target space stable and makes results comparable across settings. The learning task is formulated as image-to-gene regression that predicts spot-level expression and is intended to support biomarker discovery and tumor subtype identification.

To simulate a realistic FL scenario, we partition the dataset of 68 tissue sections into 75\% for training (51 sections) and 25\% for testing (17 sections). 
The training data is distributed such that each client receives a fixed set of 3 tissue sections. Consequently, when 17 clients participate, the entire training dataset is utilized. We adopt the pretrained ST-Net architecture based on DenseNet-121 ~\cite{he2020spatial}. ST-Net was originally trained on 30{,}612 spots from 23 patients (68 sections), where each spot’s spatially resolved expression profile is precisely matched to its corresponding H\&E image patch. This prior makes it well suited to our regression target. In each FL round, the server broadcasts the current model, clients perform one local epoch on their assigned patients, and client updates are returned and aggregated on the server according to the configured privacy mechanism.

\subsubsection{Implementation Details}
All experiments are implemented in Python using PyTorch. The HE operations use the TenSEAL library with the CKKS scheme. Because of the large number of model parameters, we set the polynomial modulus degree to 16,384 ($8192 \times 2$). Furthermore, to reduce computational load and communication overhead, we truncated the coefficient modulus bit sizes to the minimum operational requirements. 
Unless otherwise stated, runs are executed on a workstation with an Intel Core i7-13700KF CPU, 32\, GB RAM, and an NVIDIA RTX 4090 GPU (24\, GB VRAM). Preprocessing follows the ST-Net/PPML-Omics pipeline: library-size normalization per spot, log transformation of expression, extraction of 224$\times$224\, px H\&E patches centered at Visium spot coordinates, and one-time fixation of the 250 evaluation genes by global abundance ranking.

\subsubsection{Parameter Settings}
We evaluate prediction accuracy and computation time as a function of the number of clients. For a dataset with $n$ spots and $D$ target genes ($D{=}250$), the mean squared error (MSE) is defined as
\begin{equation}
\mathrm{MSE}=\frac{1}{nD}\sum_{i=1}^{n}\sum_{j=1}^{D}\bigl(y_{ij}-\hat{y}_{ij}\bigr)^{2},
\end{equation}
where $y_{ij}$ and $\hat{y}_{ij}$ denote the true and predicted expression of gene $j$ at spot $i$. A spot is the 224$\times$224\, px H\&E patch centered on a Visium spot coordinate and serves as the input unit containing local tissue morphology. Unless otherwise specified, training uses 10 global rounds with SGD (learning rate $1\times10^{-5}$, momentum $0.9$) and batch size 32, and the remaining hyperparameters are summarized in Table~\ref{tab:parameters}.

\begin{table}
\centering
\small 
\caption{Experimental Parameter Settings}
\label{tab:parameters}
\begin{tabularx}{\columnwidth}{@{} l >{\raggedright\arraybackslash}X @{}}
\hline
\textbf{Parameter} & \textbf{Value} \\
\hline \hline
\multicolumn{2}{l}{\textit{Dataset and Model}} \\
Dataset & Human Breast Cancer (23 patients / 68 sections / 30{,}612 spots) \\
Target Genes & 250 most highly expressed genes\\
Data Partitioning & 3 sections/client \\
Model Architecture & ST-Net \\
\hline
\multicolumn{2}{l}{\textit{FL Hyperparameters}} \\
Number of Clients ($N$) & 2, 5, ..., 17 \\
Number of Rounds ($T$) & 10 \\
Local Epochs ($E$) & 1 \\
Batch Size ($B$) & 32 \\
Optimizer & SGD \\
Learning Rate ($\eta$) & $1 \times 10^{-5}$ \\
\hline
\multicolumn{2}{l}{\textit{Privacy Parameters}} \\
HE Ratio ($\alpha$) & 0.2, 0.5, 0.8 \\
DP Parameters ($\epsilon,\delta$) & $(4,10^{-5}),\ (8,10^{-5})$ \\
$L_2$ Clipping Norm ($C$) & 20 \\
\hline
\multicolumn{2}{l}{\textit{HE Parameters (TenSEAL)}} \\
Polynomial Modulus Degree & 16,384 \\
Coeff. Modulus Bit Sizes & {[40, 20, 40]} \\
Global Scale & $2^{40}$ \\
\hline
\end{tabularx}
\end{table}

\subsection{Comparison Methods}
To evaluate the effectiveness of our proposed PPML-Hybrid, we compare it against two baseline methods (PPML-Omics \cite{zhou2024ppml} and HE-only), selected specifically to assess its performance in terms of MSE and computational time (FL Time).
PPML-Omics is a method that protects privacy using DP-only, and by comparing it with the proposed method, we can clearly show the impact of DP on MSE.
On the other hand, HE-only does not add noise, so there is no accuracy degradation. This allows us to evaluate the theoretical upper limit of achievable MSE. In addition, this method has very high computational costs, so it serves as a benchmark for evaluating how much our method has reduced computation time.
All results (MSE and FL Time) are reported as the average of six independent runs using different PyTorch random number seeds to control the data distribution to clients.

\subsection{Results and Discussion}
In this section, we present the experimental results to evaluate the effectiveness of our proposed PPML-Hybrid, focusing on MSE and FL Time.

\subsubsection{MSE}
\begin{figure}
\centering
\includegraphics[width=\columnwidth]{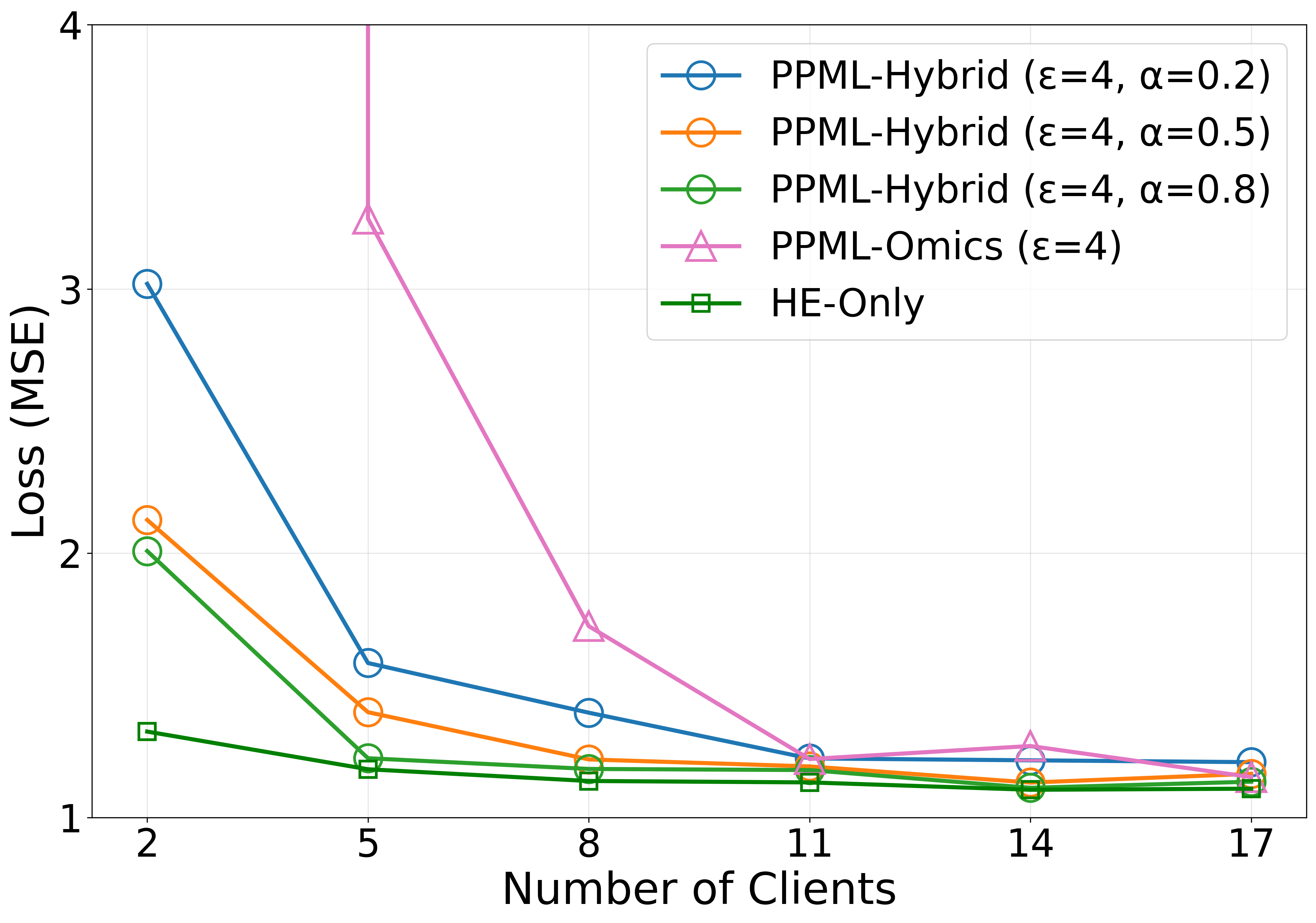}
\caption{Loss (MSE) vs. Number of Clients.}
\label{fig:mse_results}
\end{figure}
Fig.~\ref{fig:mse_results} shows the MSE of the final global model for each method as the number of clients varies from 2 to 17. A lower MSE value indicates a higher model performance. In this figure, PPML-Hybrid is plotted at $\epsilon=4$ with $\alpha\in\{0.2,0.5,0.8\}$. Note that PPML-Omics corresponds to $\alpha=0$, and HE-Only corresponds to $\alpha=1$.
As a performance upper bound, the HE-only method (green line) consistently achieves the lowest MSE across all numbers of clients, as it introduces no statistical noise. At $\epsilon=4$, our proposed PPML-Hybrid does not fully match the MSE of HE-only; however, it achieves lower MSE than PPML-Omics when the number of clients is small, and particularly with $\alpha=0.5$ and $\alpha=0.8$, it approaches the performance of HE-only. This indicates that the noise injected by the DP clients can be effectively mitigated by the noise-free updates from the HE clients in the hybrid aggregation.
In contrast, the PPML-Omics baseline (pink line, $\epsilon=4$) shows higher MSE, especially when the number of clients is small, although it improves as more clients participate. This indicates that in a realistic FL scenario, where increasing clients yields more available data, the MSE decreases with the number of clients. Overall, PPML-Hybrid approaches the non-noised HE-only baseline particularly at $\alpha=0.5$ and $\alpha=0.8$, and shows substantially greater robustness to noise than PPML-Omics at the same privacy budget.

\begin{figure}[t]
\centering
\includegraphics[width=\columnwidth]{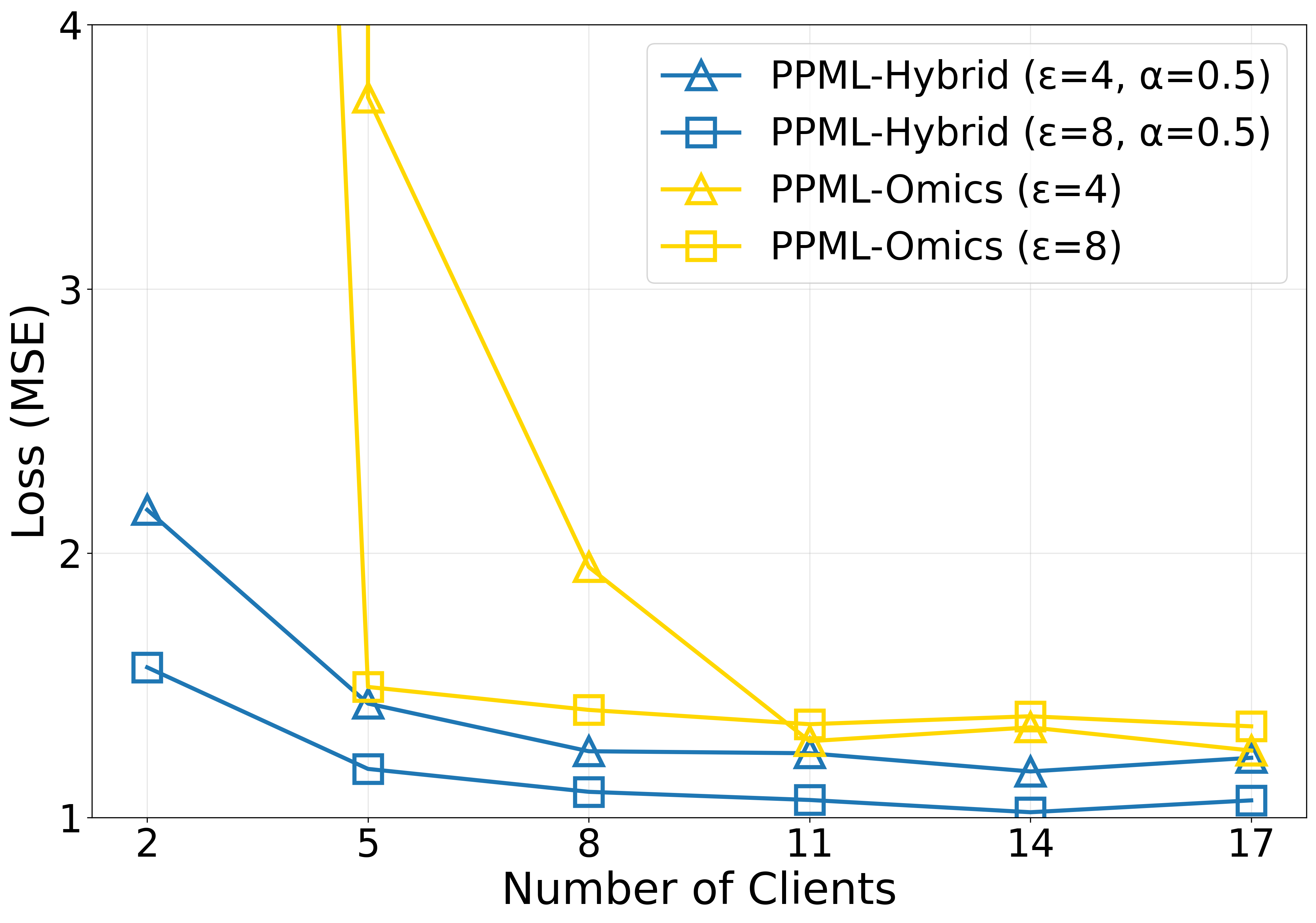}
\caption{Comparison of PPML-Hybrid and PPML-Omics at different privacy budgets ($\epsilon=4,8$) with fixed HE ratio $\alpha=0.5$.}
\label{fig:hybrid_ppmlomics_mse}
\end{figure}

Fig.~\ref{fig:hybrid_ppmlomics_mse} further compares the proposed PPML-Hybrid framework with PPML-Omics under two different privacy budgets ($\epsilon=4$ and $\epsilon=8$), while fixing the HE ratio at $\alpha=0.5$. 
As shown, both methods exhibit lower MSE as the number of clients increases, which indicates improved learning stability and data diversity. 
However, PPML-Hybrid consistently achieves lower MSE than PPML-Omics across all client settings and both privacy levels. 
This performance gain becomes particularly pronounced when the privacy budget is tighter ($\epsilon=4$), where DP-only training in PPML-Omics suffers from higher noise injection. 
At a looser privacy budget ($\epsilon=8$), both methods improve, but PPML-Hybrid still retains a clear advantage, confirming that the hybrid aggregation effectively suppresses DP-induced noise through the contribution of noise-free HE clients. 
These results validate that PPML-Hybrid not only enhances utility under strong privacy constraints but also provides stable accuracy improvements across different levels of differential privacy.

\subsubsection{Computation Time}

\begin{figure}
\centering
\includegraphics[width=\columnwidth]{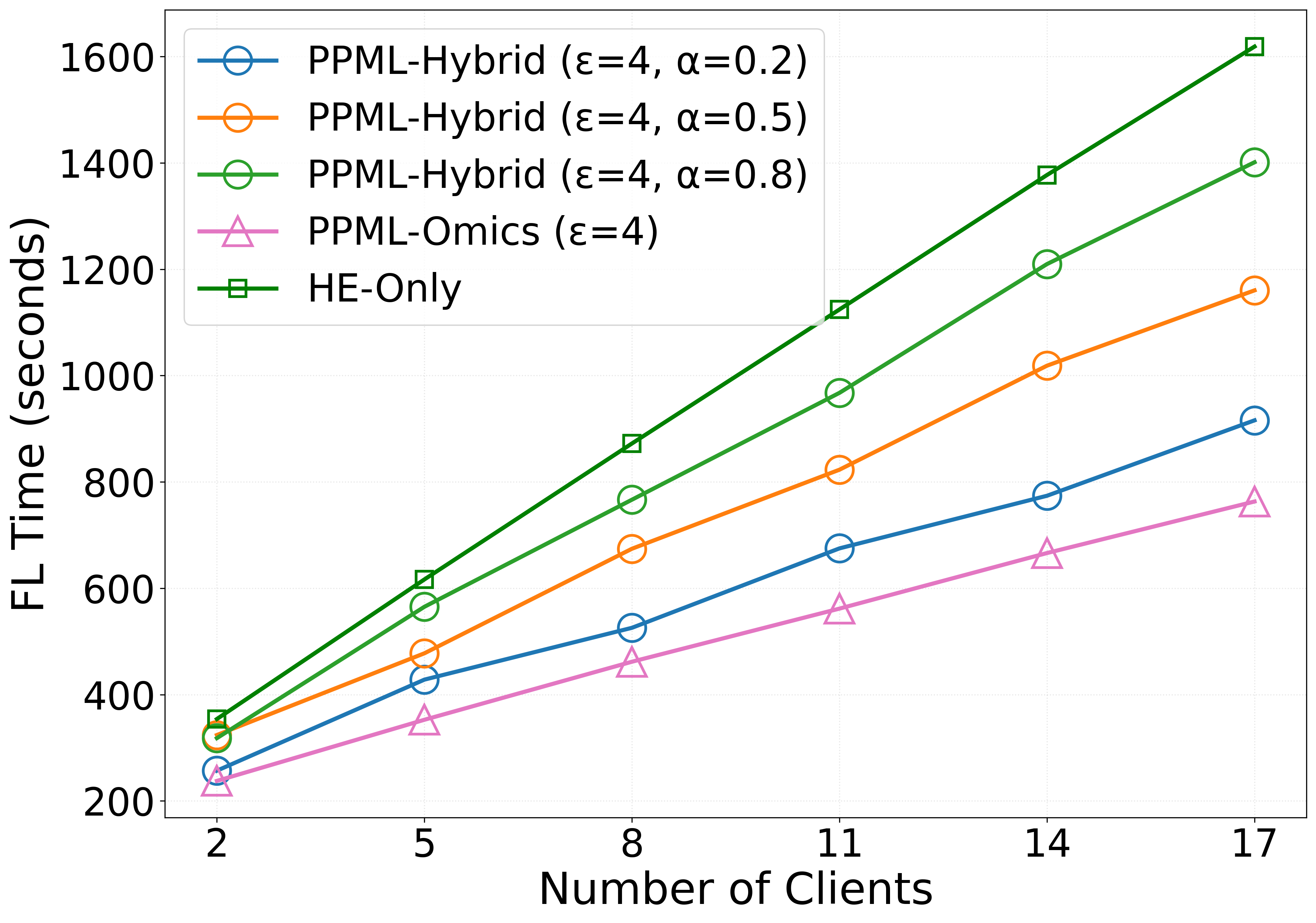}
\caption{FL Time vs. Number of Clients.}
\label{fig:time_results}
\end{figure}

Fig.~\ref{fig:time_results} compares the average computation time of each method for six runs.
Here, the computation time refers to the total time required for one FL round, 
including local model training and the corresponding privacy operations (encryption or noise addition), 
measured from the start of broadcasting the global model to the end of aggregation and testing of the global model at the server. The local model training and the corresponding privacy operations of each client are performed sequentially. 
Additionally, we assumed that the computing resources for clients within the same group (HE or DP) are equal. Note that the linear increase in computation time with respect to the number of clients is observed because the simulation executes client processes sequentially, not in parallel, for implementation convenience. Thus, the measured FL time represents the cumulative time taken by all clients to complete the learning and privacy-preserving operations before aggregation. Although the computation time in this simulation is not a result of parallel execution, in actual scenarios, execution is parallel and influenced by the slowest client in each group. Therefore, if using HE takes too long, switching to DP can effectively reduce the computation time.
In the simulation, $\epsilon$ is set to 4 while $\alpha$ is set to 0.2, 0.5, 0.8, respectively.
From Fig.~\ref{fig:time_results}, we can see that PPML-Hybrid is faster than HE-only, while slower than the PPML-Omics baseline. 
Moreover, the computation time increases as the proportion of HE group clients grows.
The reason is that, under equal computing resources, encryption processing demands more time. 
Therefore, when clients possess heterogeneous computing resources, the proposed PPML-Hybrid can reduce the overall computing time by selecting appropriate privacy methods according to the computing resources.


\section{Conclusion}
\label{sec:conclusion}

In this paper, we proposed PPML-Hybrid, a novel hybrid framework that introduces HE to FL with DP. In the proposed PPML-Hybrid, clients can select either the HE or DP method to protect privacy by itself according to their computational ability and privacy requirements. Performance evaluation results demonstrated that our proposed method maintains a high model accuracy (i.e., MSE) comparable to the HE-only baseline. Additionally, it reduces computation time per round compared to the HE-only approach. 
Moreover, given that privacy protection is theoretically achievable with either HE or DP individually, the proposed PPML-Hybrid is also capable of ensuring privacy protection by leveraging these techniques.
By offering a flexible architecture, PPML-Hybrid paves the way for the broader adoption of privacy-preserving machine learning in resource-heterogeneous environments.
In this study, specific criteria for selecting between HE and DP based on computational resources were not explicitly defined; developing concrete selection policies is left for future work. Moreover, we will extend the framework to address more realistic threat scenarios in future work.





\bibliographystyle{IEEEtran}
\bibliography{references}   

\end{document}